# AI & Data Competencies: Scaffolding holistic AI literacy in Higher Education

Kathleen Kennedy & Anuj Gupta

*Abstract*

*This chapter introduces the AI & Data Acumen Learning Outcomes Framework, a comprehensive tool designed to guide the integration of AI literacy across higher education. Developed through a collaborative process, the framework defines key AI and data-related competencies across four proficiency levels and seven knowledge dimensions. It provides a structured approach for educators to scaffold student learning in AI, balancing technical skills with ethical considerations and sociocultural awareness. The chapter outlines the framework's development process, its structure, and practical strategies for implementation in curriculum design, learning activities, and assessment. We address challenges in implementation and future directions for AI education. By offering a roadmap for developing students' holistic AI literacy, this framework prepares learners to leverage generative AI capabilities in both academic and professional contexts.*

## Introduction

**I**n the summer of 2023, a diverse task force workgroup at the University of Arizona, comprising faculty, researchers, graduate and undergraduate students, and instructional support staff, developed the AI & Data Acumen Learning Outcomes Framework. This initiative was born out of the recognition that artificial intelligence (AI) is rapidly transforming the landscape of education, research, and careers, necessitating a comprehensive approach to AI literacy in higher education. This framework is primarily intended for academic leaders, curriculum designers, and faculty members responsible for preparing students for an AI-enhanced future.

The rapid advancement and widespread adoption of AI technologies, particularly those integrating large language models (LLMs) with user-friendly interfaces such as AI chatbots, are transforming many aspects of society, including teaching and learning. These tools, exemplified by systems like ChatGPT, Claude, Gemini, and R1, as well as emerging AI agents with autonomous capabilities that can perform tasks on behalf of users, have made sophisticated AI capabilities accessible to a broad audience, creating both opportunities and challenges in educational settings.

Generative AI, in particular, represents a paradigm shift in how we teach and learn. These systems can produce human-quality text, images, and other content, opening up new possibilities for teaching and learning (Bommasani et al., 2021). For educators, this technology offers unprecedented opportunities to personalize learning, automate routine tasks, and create rich educational content. However, it also introduces significant challenges around academic integrity, information literacy, and the changing nature of core skills like writing and research. Our framework directly addresses how educators can leverage these generative capabilities while establishing appropriate guardrails to maintain educational quality and integrity.

The integration of AI into almost every profession, academic discipline, and many aspects of everyday life creates new opportunities and challenges for educators and students. In research, AI is accelerating discoveries across disciplines, from genomics to climate science, while in the professional world, AI skills are becoming increasingly crucial in STEM fields, health sciences, law, business, arts, and humanities. Preparing students for this new reality is imperative. AI literacy extends beyond technical proficiency; it encompasses building an understanding of the nature of AI, embracing human-AI co-creation, and critically thinking about AI's implications on society, ethics, and individual lives. Students must not only understand how to use AI tools but also comprehend their strengths, limitations, biases, and potential consequences. This holistic understanding is crucial for navigating this rapidly evolving technological landscape.

This chapter serves three primary purposes. First, it introduces the AI & Data Acumen Learning Outcomes Framework, a comprehensive tool designed to guide the integration of AI literacy across various disciplines in higher education. Second, we offer practical strategies using the AI & Data Acumen Learning Outcomes Framework to guide educators, curriculum committees, and academic administrators in developing courses and curriculum that will prepare students for careers and academic work where AI and data literacy is now an essential skill. Finally, we demonstrate how the framework can also be used by instructors, program directors, and assessment specialists to benchmark students' AI and data literacy progress and to identify gaps in academic programs.

The chapter is structured to provide a comprehensive exploration of the framework and its implementation by higher education professionals. We begin by discussing the rationale behind the AI & Data Acumen Learning Outcomes Framework, contextualizing it within the current landscape of AI in education. Next, we provide the theoretical background that underpins the framework, drawing from established learning theories and contemporary AI research. We then present the framework in detail, explaining its structure and the various competencies it encompasses. The chapter proceeds to offer practical guidance on applying the framework, including curriculum design strategies, learning activity development, and assessment approaches. We provide examples to illustrate the framework's application in various educational contexts, from individual course design to program-wide implementation. Finally, we discuss the challenges in implementing such a framework, including practical guidance on institutional roles and responsibilities for framework adoption, and explore future directions for AI education.

**Rationale for the AI & Data Acumen Learning Outcomes Framework**

The educational landscape is undergoing a transformation with the rapid advancement and integration of Artificial Intelligence (AI) technologies characterized by three significant developments: the advent of sophisticated generative AI in education (OpenAI, 2022), the increased adoption of AI applications in industry, services, professional fields, and research, and the integration of AI in educational processes (e.g., grading, personalized learning, administration).

The rapid integration of AI into almost every field and discipline creates a need to develop a systematic way to organize, evaluate, and manage AI and data literacy in higher education. This goes beyond technical skills, such as coding and algorithm development, to consider the broader implications and applications of AI across various disciplines. Undergraduate students now need to integrate technical skills with equally important aspects such as ethical considerations and sociocultural impacts of AI. This integration is crucial for developing a well-rounded understanding of AI (Borenstein & Howard, 2021).

**Existing AI Literacy Frameworks**

There are different approaches to systematizing AI literacy (Table 1) and existing AI literacy frameworks provide guidance for higher education contexts. That said, several of the prominent AI literacy frameworks are not easily applicable to higher education and focus on K-12 education or general AI literacy. For example, Long and Magerko's (2020) AI Literacy Framework and AI4K12's Five Big Ideas in AI (Touretzky, D., et al., 2019) are explicitly designed for K-12 students. Other AI literacy frameworks have a specific focus area (e.g., MIT AI Ethics Education Curriculum) or target industry (e.g., IEEE's Ethically Aligned Design, Google's AI Education Framework). Two prominent frameworks, UNESCO's AI Competency Framework and the EU's AI Watch and AI Alliance, though comprehensive, are not specifically tailored to higher education needs and are challenging to operationalize in higher education. This lack of focus on higher education in these frameworks represents a significant gap in addressing the unique needs and contexts of undergraduate and graduate students across various disciplines.

One AI literacy framework specifically developed for higher education is the University of Florida's (UF) AI Across the Curriculum framework (Southworth et al., 2023). This framework begins to address the unique needs of undergraduate students across all disciplines, integrating AI literacy into the university curriculum structure. While this is a very useful framework, it does not incorporate proficiency levels (such as foundational, intermediate, advanced, and expert) for each category, potentially limiting the ability to track student progress over time or provide an overarching framework for scaffolding curriculum. It also does not provide specific guidance on how AI concepts can be integrated into non-technical disciplines, explicitly address the development of students' self-efficacy in relation to AI, consider broader socio-cultural implications of AI, or address fostering innovation and creativity across disciplines.

In our work to develop the AI & Data Acumen Learning Outcomes Framework, we strived to take an integrated, culturally responsive approach to scaffolding AI literacy that explicitly recognized that AI literacy and data literacy are inextricably linked. We also set out to provide a useful tool for instructors, course developers, and students that could enable linking curriculum across disciplines and to career competencies. Finally, we set a priority on creating a framework that allows for great flexibility in how different disciplines can adapt and implement AI literacy concepts.

**Table 1:** *AI Literacy Frameworks*

| Framework Name | Author/Organization | Description | Focus Audience | Source |
| --- | --- | --- | --- | --- |
| AI Literacy Framework | Long and Magerko | Conceptual understanding of AI, focusing on mental models, strengths/limitations, and societal impact | K-12 education | Long, D., & Magerko, B., 2020 |
| AI Ethics Education Curriculum | MIT | Emphasizes ethical considerations in AI development and use, focusing on fairness, accountability, transparency, and ethics (FATE) | Higher education | MIT Media Lab, 2019 |
| Five Big Ideas in AI | AI4K12 (AAAI and CSTA) | Foundational concepts of AI: perception, representation and reasoning, learning, natural interaction, societal impact | K-12 students | Touretzky, D., et al., 2019 |
| AI Competency Framework | UNESCO | Comprehensive AI literacy covering AI and data basics, applications, ethics, governance, and sustainable development | Policymakers and educators | UNESCO, 2021 |
| AI Watch and AI Alliance | European Union | AI literacy for European citizens and workforce, covering technical skills, ethical understanding, societal impacts, and policy aspects | European citizens and workforce | European Commission, 2020 |
| Ethically Aligned Design | IEEE | Ethical considerations in AI design and development, focusing on human rights, well-being, data agency, effectiveness, transparency, and accountability | AI designers and developers | IEEE Global Initiative on Ethics of Autonomous and Intelligent Systems, 2019 |
| AI Education Framework | Google | Practical AI skills for industry applications, focusing on machine learning basics, TensorFlow programming, and AI application development | Developers and aspiring AI practitioners | Google, 2021 |

**Background**

The development of a comprehensive AI and Data Acumen Learning Outcomes Framework is grounded in established learning theories and educational principles. At the forefront are the theories of scaffolding and constructivism, which provide a solid foundation for structuring education and learning. Scaffolding, introduced by Wood, Bruner, and Ross (1976), emphasizes the importance of providing temporary support to learners as they develop new skills and knowledge, gradually removing this support as learners become more proficient. This approach aligns well with the progressive nature of AI literacy acquisition. Constructivism, as articulated by theorists like Piaget (1936) and Vygotsky (1978), posits that learners actively construct their understanding through experience and reflection, rather than passively receiving information. This theory underscores the importance of hands-on, experiential learning in AI education. Additionally, other learning theories support the development of learning competency frameworks. Bloom's Taxonomy (Bloom et al., 1956; Anderson & Krathwohl, 2001) provides a hierarchical model of cognitive processes that can be applied to structuring AI learning outcomes. Social Learning Theory (Bandura, 1977) emphasizes the importance of observation and modeling in learning, which is particularly relevant in the collaborative and rapidly evolving field of AI. Lastly, the Theory of Multiple Intelligences (Gardner, 1983) suggests that individuals have different types of intelligence, supporting the need for diverse approaches to AI education that cater to various learning styles and strengths. These theories collectively provide a robust theoretical basis for developing a multifaceted, progressive, and learner-centered AI and Data Acumen Framework.

*Interdisciplinary Approaches to AI Education*

The complex and multifaceted nature of artificial intelligence necessitates an interdisciplinary approach to AI education. This approach combines technical, social, and humanistic perspectives to provide a comprehensive understanding of AI and its implications (Eaton et al., 2022). The importance of integrating multiple disciplines in AI education stems from the recognition that AI is not merely a technical field but one that has profound social, ethical, and cultural implications (Bates et al., 2020). As Dignum (2018) argues, "AI systems are socio-technical systems and should be studied as such" (p. 1). This perspective underscores the need for AI education to go beyond programming and algorithms to include considerations of ethics, policy, and societal impact.

Combining technical and humanistic approaches allows students to develop a more nuanced understanding of AI. For instance, Grosz et al. (2019) describe a course at Harvard that integrates computer science with philosophy, providing students with both technical skills and ethical reasoning capabilities. This interdisciplinary approach enables students to critically evaluate AI systems not just for their technical performance but also for their societal implications and ethical considerations. Moreover, an interdisciplinary approach to AI education helps prepare students for the multifaceted nature of AI work in the real world. As AI systems are increasingly deployed in various sectors, professionals need to navigate complex intersections of technology, ethics, and policy (Choi et al., 2023). By exposing students to diverse perspectives and methodologies, interdisciplinary AI education equips them with the versatile skill set required in the evolving AI landscape. Zhu et al. (2021) emphasize the importance of collaborative learning environ-ments that bring together students from different disciplines to work on AI projects. Such collaborations mirror real-world scenarios and help students develop crucial skills in communication, teamwork, and interdisciplinary problem-solving.

Interdisciplinary approaches to AI education are crucial for developing well-rounded professionals capable of addressing the complex challenges posed by AI technologies. By integrating technical, social, and humanistic perspectives, these approaches prepare students for the interdisciplinary nature of AI work and foster a more comprehensive understanding of AI's role in society.

*The Need for a Structured Approach to AI Education*

The interdisciplinary nature of AI necessitates a structured approach to AI education to ensure consistency and comprehensiveness across various disciplines. As Long and Magerko (2020) argue, AI literacy encompasses a wide range of competencies, from technical skills to ethical reasoning and societal impact assessment. A structured framework helps educators address all these aspects systematically, preventing gaps in students' knowledge and skills. The multifaceted nature of AI literacy requires a holistic educational approach. Touretzky et al. (2019) propose five big ideas in AI that span perception, representation and reasoning, learning, natural interaction, and societal impact. A structured framework can ensure that all these aspects are adequately covered in AI education programs.

*The Need for a Flexible & Adaptable Framework*

The pace of AI technology and application development underscores the need for a flexible and adaptable educational framework. As Zhu et al. (2021) note, AI capabilities are evolving at an unprecedented rate, with new tools and applications emerging almost daily. This rapid evolution poses a significant challenge for educators, as traditional curricula risk becoming outdated almost as soon as they are implemented. A flexible framework, therefore, is essential to accommodate emerging technologies and shifting industry demands. Such a framework should focus on foundational concepts and adaptable skills that remain relevant despite technological changes while also providing mechanisms for the rapid integration of new developments. This approach ensures that students are equipped not just with current knowledge but with the ability to learn and adapt in an ever-changing AI landscape (Holmes et al., 2019).

*Aligning AI Education with Future Workforce Needs*

Consideration of future workforce needs is crucial for preparing students for the evolving job market. According to the World Economic Forum's Future of Jobs Report 2023, AI and machine learning specialists top the list of fastest-growing jobs, with 75% of companies planning to adopt AI technologies by 2027 (World Economic Forum, 2023). The need for data scientists and AI-savvy analysts is not new. What is new is that AI skills are increasingly becoming a top priority for employers across diverse sectors. A recent study by McKinsey (2023) found that 63% of organizations expect AI to increase their overall productivity in the next three years. This demand for AI literacy extends far beyond traditional tech roles, particularly as generative AI tools and AI agents democratize access to powerful AI capabilities across all sectors, as shown in Table 2.

**Table 2**: *AI Impact & Adoption Across Non-Technical Fields*

| Field | AI Applications | Market Indicators | Source |
|---|---|---|---|
| **Business & Management** | Decision-making, market analysis, planning, customer relationship management | 94% of business leaders agree AI is critical to success in their industry | Deloitte (2023) |
| **Healthcare** | Diagnostics, personalized treatment planning, medical image analysis | Global AI in healthcare market projected to reach $187.95 billion by 2030 | Grand View Research (2023) |
| **Legal Professions** | Legal research, contract analysis, case outcome prediction | 35% of law firms using AI for legal tasks, with adoption rates expected to rise significantly | Law.com (2023) |
| **Creative Industries** | Generative design, content creation, digital art production | AI in creative industries market expected to grow at 26.9% CAGR from 2023-2030 | Allied Market Research (2023) |
| **Education** | Personalized learning, content creation, assessment automation | 90% of countries consider AI in education a priority | UNESCO (2023) |
| **Public Policy & Governance** | Data-driven decision making, service optimization, regulatory analysis | 60% of OECD member countries have national AI strategies | OECD (2023) |

Preparing students for emerging AI-influenced careers requires a forward-looking approach to education. As Choi et al. (2023) note in their comprehensive review of AI education literature, curricula must focus not only on current tools but also on foundational concepts and adaptable skills that will remain relevant as the field advances. By adopting a structured approach to AI education that aligns with workforce needs across diverse fields, higher education institutions can better prepare students for the AI-driven future. This approach enhances students' employability and ability to contribute meaningfully to their chosen fields, regardless of whether those fields are traditionally associated with technology. The AI & Data Acumen Learning Outcomes Framework provides precisely this type of structured yet adaptable approach.

**AI & Data Acumen Learning Outcomes Framework**

The development of the AI & Data Acumen Learning Outcomes Framework was a comprehensive and meticulous process, designed to create a tool that would be both academically rigorous and practically applicable across disciplines. The process was guided by several key objectives and parameters and utilized a diverse range of information sources and analytical methods. The primary aim was to create a transdisciplinary framework that could serve multiple purposes: guide curriculum development across various academic disciplines, track students' progress in AI

& data literacy throughout their academic journey, and benchmark academic programs against a standardized set of AI & data competencies.

*AI in Higher Education & Scholarship Principles*

In developing the AI & Data Acumen Learning Outcomes Framework, we established a set of core principles to guide the use of AI in higher education and scholarship. These principles ensure that the framework aligns with fundamental academic values and pedagogical best practices, drawing from current research on ethical AI use in educational contexts (Kasneci et al., 2023; Zawacki-Richter et al., 2023).

**Table 3**: *Core Principles for AI Integration in Higher Education*

| Principle | Description | Supporting Research |
|---|---|---|
| **Human Agency** | AI should enhance human capabilities, not replace them. Educators should retain agency over core teaching decisions. | Holmes et al. (2022); Reich et al. (2023) |
| **Academic Freedom** | The choice to use or not use AI applications in a course is determined by the faculty. | AAUP (n.d.-a); AAUP (n.d.-b); Pasquale (2022) |
| **Transparency** | The use of AI should be transparent in course design, materials creation, delivery, and management. | Crawford (2021); Dignum (2022) |
| **Ethics** | AI systems should be designed and used according to principles of fairness, accountability, and mitigating bias. | Fjeld et al. (2020); UNESCO (2022) |
| **Inclusivity** | AI should be designed inclusively, avoiding marginalization of underrepresented groups. | Noble (2018); Benjamin (2019) |
| **Critical Thinking** | AI should be used to augment critical thinking, not outsource it. | Almahasees & Qadan (2023); Tang et al. (2023) |

These principles provide a foundation that strengthens the implementation of AI education and ensures that the use of AI in higher education aligns with core academic values and pedagogical best practices. As Kasneci et al. (2023) note in their comprehensive review of ChatGPT's impact on education, "maintaining human agency and critical thinking in the age of powerful AI tools is not merely advisable but essential for meaningful learning" (p. 4). Similarly, Zawacki-Richter et al. (2023) emphasize that "ethical considerations must be at the forefront of AI integration in educational settings to ensure technology serves pedagogical goals rather than undermining them" (p. 127). The AI & Data Acumen Learning Outcomes Framework operationalizes these principles through its multidimensional approach to AI literacy, ensuring that students develop not just technical skills but also the ethical reasoning, critical thinking, and sociocultural awareness needed to use AI responsibly and effectively.

*Framework Design Parameters*

To ensure the framework's effectiveness and relevance in implementing these principles, four key design parameters were established:

1. Alignment with Learning Theory: The framework needed to be grounded in established educational theories to ensure pedagogical soundness.
2. Transdisciplinary Approach: Recognizing the pervasive nature of AI across fields, the framework had to be applicable across academic disciplines.
3. Flexibility and Adaptability: Given the rapid evolution of AI technologies, the framework needed to be flexible enough to accommodate future developments.
4. Linkage to Emerging Skills: The framework had to connect directly to the most critical emerging workforce and life skills related to AI and data literacy.

By adhering to foundational principles and design parameters, the AI & Data Acumen Learning Outcomes Framework aims to provide a comprehensive and ethically grounded approach to integrating AI literacy across higher education curricula.

*Methodology*

The development process employed a multi-faceted approach, combining various sources of information and analytical methods:

- Expert Information: We consulted with a diverse group of experts from academia, industry, and policy sectors to gather insights on critical AI & data competencies.
- Statistical Analysis of Job Requirements: We conducted a comprehensive analysis of job postings across various sectors to identify the most in-demand AI and data-related skills.
- Workforce Skills Surveys: We reviewed recent workforce skills surveys to understand the current and projected needs of employers regarding AI and data literacy.
- Skills Forecasts: We incorporated data from reputable skills forecasts to ensure the framework would remain relevant in the face of rapid technological change.
- Synthesis and Iteration: The information gathered from these sources was synthesized and refined through multiple rounds of review and discussion among the task force members.

This multi-pronged approach allowed us to create a framework that is both academically rigorous and practically relevant, addressing the needs of students, educators, and employers in the rapidly evolving landscape of AI and data technologies.

**Figure 1:** *Cognitive Process Dimension Hierarchy*

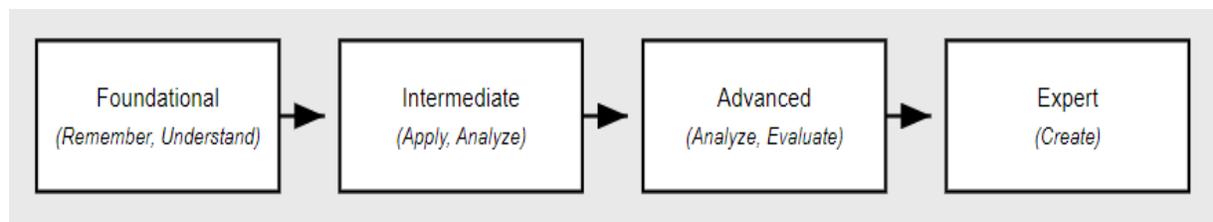

This structure aligns with contemporary understanding of cognitive development and learning processes. As Krathwohl (2002) notes, the learning hierarchy "represents a cumulative hierarchy; that is, mastery of each simpler category was a prerequisite to mastery of the next more complex one" (p. 215).

The Knowledge Dimension of the framework encompasses seven distinct but interrelated domains of knowledge and skills: Self-efficacy, Ethics, Collaboration, Socio-cultural, Innovation and Creativity, Cognitive, and Technical. This holistic approach to knowledge domains is supported by research emphasizing the importance of a holistic understanding of AI. For instance, Long and Magerko (2020) argue that AI literacy should include not only technical skills but also an understanding of AI's societal impact and ethical implications.

The inclusion of self-efficacy as a knowledge domain is particularly noteworthy. As Bandura (1997) posits, self-efficacy plays a crucial role in how people approach challenges and tasks. In the context of AI literacy, cultivating self-efficacy can empower learners to engage more confidently with AI technologies (Schunk & DiBenedetto, 2016). The ethics dimension aligns with growing recognition of the importance of ethical considerations in AI development and deployment. As Jobin et al. (2019) highlight in their global review of AI ethics guidelines, there is a growing consensus on the need for ethical AI literacy. The socio-cultural dimension acknowledges the broader context in which AI operates. This aligns with the sociotechnical perspective on AI, which emphasizes the interplay between technological systems and social contexts (Orlikowski & Scott, 2008). The innovation and creativity dimension reflects the potential of AI to augment human creativity, a concept explored by Lubart (2005) in the context of computational creativity.

By combining these cognitive processes with diverse knowledge domains, our framework provides a comprehensive roadmap for developing AI and data literacy. It recognizes that true literacy in this field requires not just technical proficiency but also critical thinking, ethical reasoning, and an understanding of broader societal implications. This two-dimensional structure allows for the creation of specific learning outcomes at the intersection of each cognitive level and knowledge domain, providing a detailed and nuanced guide for curriculum development and student assessment in AI and data literacy education.

**Table 4:** *AI & Data Acumen Learning Outcomes Framework*

| Knowledge dimensions | Competency area description | Cognitive Process Dimensions | | | |
|---|---|---|---|---|---|
| | | **Foundational** *Remember and Understand* | **Intermediate** *Apply/Use and Analyze* | **Advanced** *Analyze and Evaluate* | **Expert** *Create* |
| **Self-efficacy** | Retain and cultivate human identity, relationships, ethics, and meaning amidst increasing integration of AI in society. | Recognize aspects of one's identity and values impacted by the use of AI. | Apply strategies to maintain a strong sense of self and human relationships in an AI-infused world. | Assess the effects of AI systems on perceptions of self-worth, purpose, and autonomy. | Design plans that uphold human dignity, ethics, and community in the development and use of AI. |
| **Ethics** | Be aware and proactively address ethical AI challenges. | Recognize potential ethical issues like bias and privacy risks in AI. | Use AI in fair, transparent, and adheres to relevant norms and laws. | Evaluate solutions to mitigate ethical risks and enhance accountability in an AI application. | Develop new techniques to audit AI models, and embed ethical thinking and legal compliance within AI systems. |
| Collaboration | Facilitate greater transparency, complementarity, and reciprocity between humans and AI systems through collaborative design. | Recognize the complementary strengths of human and AI teammates. | Apply strategies for effective collaboration between cross-functional AI and non-AI experts. | Evaluate the impact of incorporating AI specialists into creative workflows and team dynamics | Create participatory systems and processes enabling reciprocal learning between humans and AI. |
| Socio-cultural | Build awareness of the historic, social, and cultural context and impact of AI. | Identify historic, social, and cultural factors that influence the use and effects of AI. | Analyze the social, and cultural effects of AI use. | Evaluate the role of historic, social, and cultural forces in the development and use of AI. | Create policy recommendations that consider social, cultural, and historic contexts. |
| Innovation and Creativity | Harness the generative capabilities of AI to augment (not replace) human creativity, communication, systems thinking, and design. | Identify how AI can enhance ideation and creative possibilities. | Apply AI techniques to expand problem-solving approaches and design options. | Evaluate improvements in systematic creativity and innovation processes enabled by AI. | Invent novel workflows integrating human creativity and AI data organization for greater impact. |
| Cognitive | Combine human cognition with the processing speed and pattern recognition of AI, while offsetting AI's biases with human judgment and context. | Describe how AI systems can complement human analytical capabilities. | Leverage AI to accelerate finding patterns and generating insights from data. | Critically evaluate AI's impact on statistical analysis and decision-making processes. | Develop new techniques combining human and artificial intelligence to enhance reasoning. |
| Technical | Develop expertise to ethically and responsibly build, apply, and advance AI systems and tools. | Define basic concepts of major AI approaches like machine learning, neural networks, and deep learning. | Use common AI algorithms for data analysis tasks. | Evaluate tradeoffs in AI model selection, optimization, and production deployment. | Advance the state-of-the-art in AI algorithms to expand capabilities and mitigate risks (e.g., type I/type II errors, security risk). |

This framework serves as a practical tool for educators, offering a structured approach to scaffold student learning in AI and data literacy. It provides concrete guidance for integrating AI concepts across various disciplines and levels of education, helping teachers to systematically develop students' AI competencies.

## Applying the AI & Data Acumen Learning Outcomes Framework

The AI & Data Acumen Learning Outcomes Framework provides a roadmap for integrating AI literacy across higher education. This section outlines strategies for implementing the framework in curriculum design, learning activity development, assessment, and faculty support. A key strength of the AI & Data Acumen Learning Outcomes Framework is its balanced approach to cultivating both career-relevant skills and academic capabilities. While preparing students for the AI-driven job market, it also emphasizes the development of critical thinking, ethical reasoning, and research skills essential for academic pursuits in AI-related fields.

**Curriculum Design and Alignment**

To effectively embed AI literacy across various subjects, institutions should adopt a transdisciplinary approach. This involves identifying points of intersection between AI and existing disciplinary content and developing strategies to seamlessly integrate AI concepts (Zhu et al., 2021). Strategies for embedding AI topics in various subjects include:

- Incorporating AI-related case studies and examples in existing courses.
- Developing interdisciplinary modules that explore AI applications in specific fields.
- Creating collaborative projects that require students to apply AI concepts to discipline-specific problems.

An Example of a Cross-Disciplinary AI Project: a collaboration between computer science and sociology departments could involve students developing an AI algorithm to analyze social media data for trends in public opinion on a current issue. This project would integrate technical skills with sociocultural awareness, aligning with multiple dimensions of the framework.

**Designing New AI-Focused Courses and Programs**

When creating new AI-focused curricula, it is crucial to balance technical skills with broader competencies outlined in the framework. Guidelines for creating comprehensive AI curricula include:

- Ensure coverage of all seven knowledge dimensions: Self-efficacy, Ethics, Collaboration, Socio-cultural, Innovation and Creativity, Cognitive, Technical.
- Progressively build competencies from foundational to expert levels.
- Incorporate hands-on, project-based learning experiences.
- Integrate ethical considerations and societal impacts throughout the curriculum.

To balance technical skills with broader competencies, curricula should include courses that focus on the ethical, sociocultural, and collaborative aspects of AI alongside technical courses. For example, a course on "AI Ethics and Society" could be a required component of an AI-focused program (Grosz et al., 2019).

*Learning Activity Development*

The framework can guide learning activity development. Hands-on projects, case studies, and ethical dilemmas are effective ways to engage students with the multifaceted nature of AI literacy. Some examples include:

- Technical Dimension: Develop a machine learning model to predict housing prices, focusing on data preprocessing, model selection, and evaluation.
- Ethical Dimension: Analyze a real-world case of AI bias (e.g., facial recognition systems) and propose mitigation strategies.
- Socio-cultural Dimension: Conduct a research project on the impact of AI on a specific industry or community, exploring both benefits and challenges.

Collaborative AI development exercises could involve:

- Group projects to develop AI-powered solutions for local community issues.
- Hackathons that bring together students from various disciplines to solve AI-related challenges.
- Role-playing exercises where students take on different stakeholder perspectives in AI development scenarios.

*Balancing Technical Skills with Ethical & Sociocultural Awareness*

To integrate ethical considerations into technical projects, consider:

- Requiring students to conduct an ethical impact assessment as part of their technical development process
- Incorporating discussions on potential biases and societal implications at each stage of AI project development
- Encouraging students to develop guidelines for responsible AI use alongside their technical solutions

Activities that explore AI's societal impacts could include:

- Debates on controversial AI applications (e.g., autonomous weapons, predictive policing)
- Scenario planning exercises to anticipate future societal changes due to AI advancements
- Guest lectures from industry professionals and ethicists to provide real-world perspectives

*Skills Assessment Strategies*

Continuous evaluation of AI competencies is crucial for tracking student progress across the framework's dimensions. Formative assessment strategies could include:

- Regular quizzes on AI concepts and their applications
- Peer reviews of AI project proposals and implementations
- Reflective journals documenting students' evolving understanding of AI's implications

For summative assessments, design approaches that reflect real-world AI challenges:

- Capstone projects that require students to develop and present an AI solution, including ethical considerations and societal impact analysis
- Comprehensive exams that test both technical knowledge and broader AI literacy concepts
- Industry-partnered projects where students solve actual AI-related problems faced by organizations

**Portfolio-Based Assessment for AI Competencies:** Encouraging students to document their AI journey through portfolios can provide a holistic view of their developing competencies. Portfolios could include:

- Code repositories showcasing technical projects
- Written reflections on ethical dilemmas encountered in AI development
- Documentation of collaborative AI projects, highlighting teamwork and communication skills

**Showcases:** To showcase interdisciplinary AI projects, institutions could:

- Organize AI project displays or symposiums where students present their work to peers, faculty, and industry representatives
- Create online platforms for students to share their AI portfolios, fostering a community of practice
- Encourage students to participate in AI competitions or conferences to gain external validation of their skills

*Faculty Development & Support*

Faculty development and support are crucial components in successfully implementing the AI & Data Acumen Learning Outcomes Framework. To begin, institutions should assess faculty AI literacy needs through a multi-faceted approach. This could include conducting surveys to gauge current AI knowledge and teaching practices, analyzing course syllabi to identify gaps in AI coverage across disciplines, and hosting faculty focus groups to discuss challenges in integrating AI into teaching. Based on these assessments, targeted training programs can be created to address specific needs. These might include workshops on specific AI technologies and their applications in various fields, seminars on ethical considerations in AI led by experts, and hands-on training sessions for using AI tools in teaching and research.

*Creating Resources for AI-Enhanced Teaching*

Creating resources for AI-enhanced teaching is also essential. Institutions should develop comprehensive guides for integrating AI tools in teaching, which could cover best practices for using AI-powered educational technologies, strategies for incorporating AI concepts into existing course content, and guidelines for designing AI-related assignments and projects. To foster ongoing learning and collaboration, establishing communities of practice for AI education is highly beneficial. This can be achieved by creating online forums or discussion groups for faculty to share experiences and resources, organizing regular meetups or brown bag sessions focused on AI in education, and facilitating mentorship programs that pair AI-experienced faculty with those new to the field. These initiatives collectively support faculty in developing the skills and confidence needed to effectively integrate AI literacy across the curriculum.

*Institutional Benchmarking Tool*

The AI & Data Acumen Learning Outcomes Framework can serve as a benchmarking tool for institutions to assess their AI education efforts. By mapping existing courses and programs to the framework's dimensions and proficiency levels, institutions can identify strengths and areas for improvement in their AI literacy initiatives.

To implement the framework as a benchmarking tool:

1. Create a matrix aligning courses with framework dimensions and proficiency levels
2. Conduct regular audits of AI-related content across disciplines
3. Set institutional goals for AI literacy coverage and track progress over time
4. Use benchmarking results to inform curriculum development and resource allocation

By systematically applying the AI & Data Acumen Learning Outcomes Framework across curriculum design, learning activities, assessment strategies, and faculty development, institutions can create a comprehensive and effective approach to AI literacy education. This holistic implementation ensures that students develop the multifaceted competencies necessary to navigate the AI-driven future confidently and ethically.

*Career Preparation & Readiness Benchmarking Tool*

The AI & Data Acumen Learning Outcomes Framework can offer a systematic way to align educational outcomes with industry needs. Effective implementation requires coordinated institutional efforts with defined responsibilities. Academic leadership establishes governance structures for framework adoption, while curriculum committees evaluate job requirements across the seven knowledge dimensions, assigning appropriate proficiency levels. Career services analyzes job specifications to create AI competency profiles, which academic advisors then use to guide students in evaluating their progress against career requirements. Assessment specialists gather data to identify programmatic strengths and gaps, enabling department chairs to evaluate their AI education effectiveness. This approach ensures continuous curriculum refinement aligned with evolving industry demands, empowering students to make informed decisions about their learning pathways in AI while providing clear institutional guidance.

## Challenges & Future Directions

The implementation of the AI & Data Acumen Learning Outcomes Framework, while promising, is not without its challenges. One of the primary barriers institutions may face is resource constraints, particularly in terms of funding and expertise. Many institutions may struggle to allocate sufficient resources for faculty training, resource development, infrastructure upgrades, and curriculum redesign necessary for comprehensive AI education. Additionally, there may be resistance to change from faculty members accustomed to traditional curricula, necessitating careful change management strategies. Overcoming these barriers will require strong institutional commitment, strategic resource allocation, and ongoing support for faculty development.

The rapid evolution of AI technologies presents another significant challenge. As AI capabilities expand and new applications emerge, maintaining the relevance of the framework becomes crucial. Institutions must develop strategies to keep the framework up-to-date, which may include establishing partnerships with industry leaders, participating in AI research communities, and creating mechanisms for rapid curriculum updates. Regular revision cycles should be built into the framework's implementation plan, ensuring that learning outcomes and competencies remain aligned with the latest developments in AI. Particularly with generative AI tools rapidly transforming educational practices, the framework must evolve to address these emerging capabilities and their implications.

Ensuring inclusivity and accessibility in AI education is paramount as the field grows. There is a risk that AI education may inadvertently perpetuate existing biases or create new ones, particularly if the diversity of AI's impacts across different communities is not adequately addressed. Institutions must actively work to make AI literacy accessible to diverse learner populations, considering factors such as socioeconomic background, prior technical experience, and learning styles. This may involve developing adaptive learning pathways, providing additional support resources, and incorporating diverse perspectives and case studies into the curriculum.

Implementation responsibility should be clearly defined within institutional structures, with academic leadership, curriculum committees, and faculty development offices collaborating to integrate the framework effectively. Specific roles should include framework champions within each academic department, dedicated support staff for faculty development, and assessment specialists to evaluate implementation effectiveness.

The ongoing refinement and validation of the framework will be critical to its long-term success. Institutions should establish robust mechanisms for collecting feedback from students, faculty, industry partners, and other stakeholders. This feedback should be systematically analyzed and incorporated into regular framework updates. Future research directions should focus on validating the framework's effectiveness in preparing students for AI-related careers and fostering critical AI literacy. This may involve longitudinal studies tracking student outcomes, comparative analyses of different implementation approaches, and assessments of the framework's impact on broader societal understanding of AI.

Addressing these challenges and pursuing these future directions will require collaborative efforts across academia, industry, and policy-making bodies. By anticipating and proactively addressing these issues, institutions can ensure that the AI & Data Acumen Learning Outcomes Framework remains a dynamic and effective tool for preparing students to navigate the AI-driven future. As AI continues to transform various aspects of society, the ability to adapt and evolve this educational framework will be crucial in empowering the next generation with the skills and knowledge needed to shape the ethical and responsible development of AI technologies.

**Conclusion**

The AI & Data Acumen Learning Outcomes Framework marks a significant advancement in addressing the need for comprehensive AI literacy in higher education. By defining seven key knowledge dimensions across four proficiency levels, the framework offers a holistic approach to AI education. Its strength lies in its comprehensiveness, addressing not only technical skills but also critical aspects such as ethical considerations, societal impacts, and human-AI collaboration. The framework's adaptability allows for implementation across various educational contexts, ensuring its relevance in diverse academic settings.

As AI continues to rapidly transform industries, society, and educational practices, this framework serves as a vital roadmap for AI education. Its comprehensive approach to integrating generative AI literacy across the curriculum directly addresses the question of how educators can leverage AI's expanding capabilities to improve teaching and learning while minimizing educational risks. By providing structured guidance for curriculum development, learning activities, and assessment, the framework empowers educational institutions to prepare students for meaningful human-AI collaboration in their academic and professional futures.

The framework's potential impact extends beyond individual courses to reshaping entire higher education curricula. By aligning educational outcomes with the multifaceted nature of AI, it equips students with the technical proficiency, ethical reasoning, and sociocultural awareness necessary to navigate complex AI challenges. This balanced approach ensures graduates can not only use AI tools effectively but also critically evaluate their implications and guide their responsible development.

We encourage academic leaders, curriculum designers, and faculty members to adopt and adapt this framework to their specific contexts. Its success and ongoing relevance depend on widespread implementation and continuous refinement based on real-world experiences. We particularly urge institutions to establish clear governance structures for implementation, with defined roles and responsibilities for academic departments, curriculum committees, and faculty development offices. This collaborative approach will ensure the framework becomes an integral part of institutional strategy rather than an isolated initiative.

We invite collaboration from across the educational spectrum to further develop and enhance this framework. By sharing insights, best practices, and outcomes, we can collectively advance AI education and ensure that future generations are well-prepared to lead in an AI-infused world. Through this collective effort, we can address the critical need for an AI-literate society that can harness the benefits of these powerful technologies while mitigating their risks.

The AI & Data Acumen Learning Outcomes Framework provides a solid foundation for elevating AI literacy in higher education. As generative AI and other technologies continue to evolve, this framework offers a flexible yet structured approach for empowering students with the comprehensive skills and ethical grounding needed to thrive in our AI-driven future. By embracing and implementing this framework, educational institutions can prepare their students for success and contribute to the responsible advancement of AI.